\documentclass[prd,aps,showpacs,tightenlines]{revtex4}  
\usepackage{mathrsfs}
\usepackage{amsmath}
\usepackage{amssymb}
\usepackage{epsfig}
\usepackage{graphicx}
\usepackage{booktabs}
\usepackage{multirow}
\usepackage{subfigure}
\begin{document}
\newcommand{\psl}{ p \hspace{-1.8truemm}/ }
\newcommand{\nsl}{ n \hspace{-2.2truemm}/ }
\newcommand{\vsl}{ v \hspace{-2.2truemm}/ }
\newcommand{\epsl}{\epsilon \hspace{-1.8truemm}/\,  }

\title{Semileptonic decays of $B_c$ meson to $P$-wave charmonium states }
\author{Zhou Rui$^1$}\email{jindui1127@126.com}
\author{Jie Zhang$^1$}
\author{Li-li Zhang$^2$}
\affiliation{$^1$College of Sciences, North China University of Science and Technology,
                          Tangshan 063009,  China}
\affiliation{$^2$Center for Publishing, North China University of Science and Technology,
                          Tangshan 063009,  China}
\date{\today}
\begin{abstract}
Inspired by a series of unexpected measurements of  semileptonic
decays mediated via $b\rightarrow c $ charged current interactions,
we explore semileptonic $B_c$ decays to the four lightest $P$-wave charmonium states, $\chi_{c0}, \chi_{c1}, \chi_{c2}, h_c$,
by the recently developed improved perturbative QCD formalism,
in which the charm quark mass effect is included  both in the Sudakov factor and the hard kernels.
We first directly evaluate the concerned
transition form factors  with vector and  axial-vector currents
in the region of small momentum transfer squared
and then recast them to the full  kinematical region by adopting the exponential parametrization.
The obtained   form factors are used to evaluate the semileptonic decay branching ratios, which can reach the order of  $10^{-3}$,
letting the corresponding measurement appear feasible.
For a better analysis, a comparison of our results with the predictions of other models is provided.
We also present  the ratios between the tau and light lepton branching ratios
 and the polarization contributions  in the relevant  processes,
which still need experimental tests in the ongoing and forthcoming experiments.
Any significant deviations from the Standard Model results may provide some hints of new physics effects.
\end{abstract}



\maketitle

\section{Introduction}
Recently, a number of experimental measurements involving semitauonic  decays of the
charged current $b\rightarrow c \tau \nu_{\tau}$ transitions
 have shown interesting deviations from their Standard Model (SM) expectations,
  though the significance of the excess is low due to the large statistical uncertainties.
 For example, the measured values for  $\mathcal{R}(D^{(*)}) $ corresponding to the ratios of branching fractions
$\mathcal{B}(B\rightarrow D^{(*)} \tau \bar{\nu}_{\tau})/\mathcal{B}(B\rightarrow D^{(*)} l\bar{\nu}_{l})$,
with $l$ either an electron or  muon,
 by the $BABAR$ \cite{prl109101802,prd88072012}, Belle \cite{prd92072014,prd94072007,prl118211801,prd97012004},
 and LHCb \cite{prl115111803,170808856,171102505} Collaborations show a significant excess over the SM expectation
  \cite{prd85094025,prd92034506,prd92054510,HFLAV2016}.
The most statistically significant deviation at the $4\sigma$  level \cite{HFLAV2016}
is seen in the combination of $\mathcal{R}(D) $ and $\mathcal{R}(D^*)$.
Very recently, 
the corresponding measurement regarding $b\rightarrow c \tau \nu_{\tau}$ in $B_c$
had also been reported by LHCb \cite{prl120121801}
\begin{eqnarray}\label{eq:dfenzhibi}
 \mathcal{R}(J/\psi)=\frac{\mathcal{B}(B^+_c\rightarrow J/\psi \tau^+ \nu_{\tau})}{\mathcal{B}(B^+_c\rightarrow J/\psi \mu^+ \nu_{\mu})}=
 0.71\pm0.17(\text{stat})\pm0.18(\text{syst}).
\end{eqnarray}
The yield value lies at about $2\sigma$ above the range of existing predictions in the SM \cite{epjc76564,cpc37093102,prd87014009}.
These ratios have been calculated to high precision due to the
cancellation of numerous uncertainties common to the numerator and  denominator.
Within the SM, the deviation from  unity is mainly caused by the massive $\tau$ lepton,
which also increases the sensitivity to new physics (NP) in these decays.
Then, the possible NP effects in the semileptonic decays   have been discussed recently in several papers \cite{prd97054014,plb7765,plb77952,prd96076001,170909989,171000351,171004127,180100917,180103375}.
To maximize future sensitivity to NP contributions, measuring and understanding
the semileptonic modes involving various
$P$-wave orbitally excited charmonium $X(X\in \{ \chi_{c0},\chi_{c1},\chi_{c2},h_c\})$ in the final state for the same flavor content
are important and necessary,
not only as they can give additional and complementary information on the NP
but also as they constitute backgrounds to the $\mathcal{R}(J/\psi)$ measurements.

Experimentally, many nonleptonic decays with $J/\psi$ or $\psi(2S)$ as the final charmonium have been detected \cite{pdg2016},
and the first evidence for the decay $B_c\rightarrow \chi_{c0}\pi$ is found at $4.0\sigma$ significance by the LHCb experiment \cite{prd94091102}.
However, for the semileptonic decays, so far,  only the $B_c\rightarrow J/\psi $ transitions
 have recently been observed by the LHCb Collaboration \cite{prl120121801,prd90032009}.
As the LHC accumulates more and more data, the semileptonic $B_c$ decays to the $P$-wave charmonium
 will have more possibilities to be detected. Theoretically,
 essential to the study of the  semileptonic decays is the
calculation of the invariant form factors describing the corresponding hadronic transitions.
In the literature, a wide range of various  approaches has been used to compute the $B_c\rightarrow X$   transition form factors,
such as the  QCD sum rules (QCDSR) \cite{prd77054003,prd79116001},
the covariant light-front  quark model (LFQM) \cite{prd79114018}, the renormalization group method (RGM) \cite{prd65014017},
the relativistic constituent quark model (RCQM) \cite{prd73054024}, relativistic quark model (RQM) \cite{prd71094006},
the nonrelativistic quark model (NRQM) \cite{prd74074008},
the Bethe-Salpeter approach (BS) \cite{jpg39015009},
the  relativistic quark model based on the quasipotential approach (RQMQP) \cite{prd82034019},
and the Isgur-Scora-Grinstein-Wise II model (ISGW II) \cite{prd87034004}.
More recently, the relativistic corrections to the form factors of
the $B_c$ meson into $P$-wave orbitally excited charmonium
have been investigated using the nonrelativistic QCD effective theory (NRQCD) \cite{171007011}.

As a successive work of \cite{epjc76564,prd97033001,epjc78463}, in this paper,
 we do not attempt to resolve the $\mathcal{R}(J/\psi)$ anomaly beyond the SM, but
 provide more reliable calculations of those orbitally  excited state modes within the SM.
A future improvable measurement might reveal whether a similar anomaly also exists in  $\mathcal{R}(X)$.
In order to meet the measurements for charmonium $B_c$ decays with good precision,
we adopt the so-called improved perturbative QCD formalism  \cite{180106145} recently developed by Xin Liu {\it et al.}.
The charmonium $B_c$ decays are a multiscale process, 
which contain three scales: the bottom quark mass $m_b$, the charm quark mass $m_c$, and the QCD scale $\Lambda_{QCD}$.
Under the hierarchy of $m_b\gg m_c\gg \Lambda_{QCD}$, the charm quark effect enters the  Sudakov exponent through
an additional large  infrared logarithm $\log{(m_b/m_c)}$, which should   be resummed.
For the detailed derivation of the $k_T$ resummation technique with the finite charm quark mass, the reader is referred to \cite{180106145}.

The outline of the paper is as follows: In Sec. \ref{sec:framework},
we define kinematics and describe the meson distribution amplitude  of the initial and final states.
In Sec. \ref{sec:formfactor}, we give the factorization formulas for the $B_c\rightarrow X$ form factors in the PQCD approach.
Subsequently, we present the general formalism for the semileptonic differential decay widths
with the lepton-helicity states.
Section. \ref{sec:results} is devoted to the numerical analysis 
of the form factors,   branching ratios,  polarizations   and comparison of our results with the other approaches.
 A summary is given in Sec. \ref{sec:sum}.


\section{ KINEMATICS and meson distribution amplitudes   }\label{sec:framework}
\begin{figure}[tbp]
\begin{center}
\centerline{\epsfxsize=7 cm \epsffile{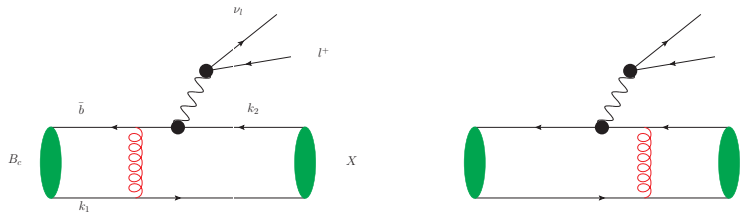}}
\vspace{1.6cm} \caption{The leading-order Feynman diagrams for the
 semileptonic decays  $B^+_c\rightarrow X l^+ \nu_l$ with $l=(e,\mu,\tau)$.}
 \label{fig:semi}
 \end{center}
\end{figure}

For simplicity we  work in the rest frame of the $B_c$ meson and use light-cone coordinates.
The momentum of the $B_c$ meson and charmonium can be denoted as \cite{epjc76564,prd67054028,cpc37093102}
\begin{eqnarray}\label{eq:def}
P_1=\frac{M}{\sqrt{2}}(1,1,\textbf{0}_{\rm T}),\quad P_2=\frac{M}{\sqrt{2}}(r\eta^+,r\eta^-,\textbf{0}_{\rm T}),
\end{eqnarray}
with the ratio $r=m/M $, and $m(M)$ is the mass of the charmonium ($B_c$) meson.
The factors  $\eta^{\pm}=\eta\pm \sqrt{\eta^2-1}$ 
are defined in terms of the velocity transfer $\eta=v_1 \cdot v_2$ with $v_1=P_1/M$ and $v_2=P_2/m$ \cite{prd67054028}.
For the momentum transfer $q=P_1-P_2$, there exists $\eta=\frac{1+r^2}{2r}-\frac{q^2}{2rM^2}$.
The momentum of the valence quarks $k_{1,2}$, whose notation are displayed in Fig \ref{fig:semi},
are parametrized as
\begin{eqnarray}
k_1=x_1P_1+(0,0,\textbf{k}_{\rm {1T}}), \quad k_2=x_2P_2+(0,0,\textbf{k}_{\rm {2T}}),
\end{eqnarray}
where the $\textbf{k}_{\rm{1T,2T}}$, $x_{1,2}$ represent the transverse
momentum and longitudinal  momentum fraction of the charm quark inside the meson, respectively.

As the direct analogue of the vector charmonium  \cite{epjc76564}, for an axial-vector charmonium,
 the longitudinal (transverse) polarization vectors $\epsilon(0(\pm))$ can be defined as
 \begin{eqnarray}\label{eq:polar}
\epsilon(0)&=&\frac{1}{\sqrt{2}} (\eta^+,-\eta^-,\textbf{0}_{\rm T}), \quad \epsilon(\pm)=(0,0,\textbf{1}_{\rm T}),
\end{eqnarray}
which satisfy the normalization $\epsilon^2(0)= \epsilon^2(\pm)=-1$
and the orthogonality $\epsilon(0) \cdot P_2=0$.

For the tensor charmonium, since the polarization tensor $\epsilon_{\mu\nu}(\lambda)$ with helicity $\lambda$ is traceless,
symmetric and satisfies the condition $\epsilon_{\mu\nu}(\lambda)P_2^{\nu}=0$,
it can be constructed via the polarization vector $\epsilon(0,\pm)$  \cite{prd82054019,prd83034001}:
\begin{eqnarray}\label{eq:tensor}
\epsilon_{\mu\nu}(\pm 2)&=&\epsilon_{\mu}(\pm)\epsilon_{\nu}(\pm), \nonumber\\
\epsilon_{\mu\nu}(\pm 1)&=&\frac{1}{\sqrt{2}}[\epsilon_{\mu}(\pm)\epsilon_{\nu}(0)+\epsilon_{\nu}(\pm)\epsilon_{\mu}(0)], \nonumber\\
\epsilon_{\mu\nu}(0)&=&\frac{1}{\sqrt{6}}[\epsilon_{\mu}(+)\epsilon_{\nu}(-)+\epsilon_{\mu}(-)\epsilon_{\nu}(+)]+
\sqrt{\frac{2}{3}}\epsilon_{\mu}(0)\epsilon_{\nu}(0).
\end{eqnarray}
As usual, it is convenient to define another two polarization vectors $\epsilon_{T\mu}$ and $\epsilon_{\bullet \mu}$
corresponding to the  transition form factors and light-cone distribution amplitudes (LCDAs), respectively,
which are related to the polarization tensor by \cite{prd83014008}
\begin{eqnarray}\label{eq:ano}
\epsilon_{T\mu}(\lambda)=\frac{\epsilon_{\mu\nu}(\lambda)P_1^{\nu}}{M}, \quad
\epsilon_{\bullet \mu}(\lambda)=m\frac{\epsilon_{\mu\nu}(\lambda)v^{\nu}}{P_2\cdot v},
\end{eqnarray}
with the unit vectors $v=(0,1,\textbf{0}_{\rm T})$ on the light cone.
Combining Eqs.(\ref{eq:def}), and (\ref{eq:polar})-(\ref{eq:ano}), we further have
\begin{eqnarray}\label{eq:vvv}
\epsilon_{T \mu}(\pm 2)&=&0,\nonumber\\
\epsilon_{T \mu}(\pm 1)&=& \sqrt{\frac{1}{2}}\frac{\epsilon(0)\cdot P_1}{M}\epsilon_{\mu}(\pm)= \sqrt{\frac{1}{2}}\sqrt{\eta^2-1}\epsilon_{\mu}(\pm),\nonumber\\
\epsilon_{T \mu}(0)&=& \sqrt{\frac{2}{3}}\frac{\epsilon(0)\cdot P_1}{M}\epsilon_{\mu}(0)=
\sqrt{\frac{2}{3}}\sqrt{\eta^2-1}\epsilon_{\mu}(0),\nonumber\\
\epsilon_{\bullet \mu}(\lambda)&=&\frac{\epsilon_{T \mu}(\lambda)}{\sqrt{\eta^2-1}}.
\end{eqnarray}
Note that both $\epsilon_{T}$ and  $\epsilon_{\bullet}$ above have the same energy scaling as the
usual polarization vector $\epsilon$. It makes the calculations of the $B_c$ decays into a tensor meson 
similar  to those of the vector analogues. 
The only difference is that the polarization vector $\epsilon$ is replaced by $\epsilon_{\bullet}$
 in the LCDAs but by $\epsilon_{T}$ in the transition form factors.

 In the course of the PQCD calculations,  the necessary inputs contain
the LCDAs, which are constructed via the nonlocal matrix elements.
The $B_c$ meson is a heavy-light system, whose light-cone matrix element can
be decomposed as
\begin{eqnarray}
\int d^4z e^{ik_1\cdot z}\langle 0|\bar{b}_{\alpha}(0)c_{\beta}(z)|B_c(P_1)\rangle=
\frac{i}{\sqrt{2N_c}}[(P_1+M)\gamma_5\phi_{B_c}(k_1)]_{\beta\alpha},
\end{eqnarray}
where $N_c=3$  is the color factor. Here, we only consider one of the dominant Lorentz structures.
In coordinate space the distribution amplitude $\phi_{B_c}$ with an intrinsic $b$
(the conjugate space coordinate to $k_T$) dependence
is adopted in a Gaussian form as \cite{180106145}
\begin{eqnarray}
\phi_{B_c}(x,b)=N_{B_c}x(1-x)\exp{[-\frac{(1-x)m_c^2+xm_b^2}{8\omega^2x(1-x)}-2\omega^2b^2x(1-x)]},
\end{eqnarray}
with the shape parameter $\omega=1.0\pm 0.1$ GeV related to the factor $N_{B_c}$ by the normalization
 \begin{eqnarray}
\int^1_0\phi_{B_c}(x,0)dx=1.
\end{eqnarray}

For the $P$-wave charmonium states, their LCDAs were recently analyzed in Ref. \cite{prd97033001}
and  are defined by
\begin{eqnarray}\label{eq:non}
\langle S (P_2)|\bar{c}_{\alpha}(z)c_{\beta}(0)|0\rangle &=&\frac{1}{\sqrt{2N_c}}\int_0^1dx e^{ixP_2\cdot z}
[\rlap{/}{P}_2\psi_S^v(x)+m\psi_S^s(x)]_{\beta\alpha}.\nonumber\\
\langle A (P_2,\epsilon(0))|\bar{c}_{\alpha}(z)c_{\beta}(0)|0\rangle &=&\frac{1}{\sqrt{2N_c}}\int_0^1dx e^{ixP_2\cdot z}
[m\gamma_5\rlap{/}{\epsilon}(0) {\psi^L_A(x)}+\gamma_5\rlap{/}{\epsilon}(0)\rlap{/}{P}_2 \psi_A^t(x)]_{\beta\alpha},\nonumber\\
\langle A (P_2,\epsilon(\pm))|\bar{c}_{\alpha}(z)c_{\beta}(0)|0\rangle &=&\frac{1}{\sqrt{2N_c}}\int_0^1dx e^{ixP_2\cdot z}
[m\gamma_5\rlap{/}{\epsilon}(\pm) { \psi_A^V(x)}+\gamma_5\rlap{/}{\epsilon}(\pm)\rlap{/}{P}_2\psi_A^T(x)]_{\beta\alpha},\nonumber\\
\langle T (P_2,\epsilon_{\bullet}(0))|\bar{c}_{\alpha}(z)c_{\beta}(0)|0\rangle &=&\frac{1}{\sqrt{2N_c}}\int_0^1dx e^{ixP_2\cdot z}
[m\rlap{/}{\epsilon}_{\bullet}(0) \psi_T(x)+\rlap{/}{\epsilon}_{\bullet }(0)\rlap{/}{P}_2 \psi^t_T(x)]_{\beta\alpha},\nonumber\\
\langle T (P_2,\epsilon_{\bullet}(\pm))|\bar{c}_{\alpha}(z)c_{\beta}(0)|0\rangle &=&\frac{1}{\sqrt{2N_c}}\int_0^1dx e^{ixP_2\cdot z}
[m\rlap{/}{\epsilon}_{\bullet }(\pm)  \psi^V_T(x)+\rlap{/}{\epsilon}_{\bullet}(\pm)\rlap{/}{P}_2 \psi^T_T(x)]_{\beta\alpha},
\end{eqnarray}
where the abbreviations $S$, $A$, and $T$ correspond to scalar, axial-vector, and
tensor charmonium states, respectively. $\psi_S^v$, $\psi^{L,T}_A$, and $\psi^{(T)}_T$ are of  twist-2,
while $\psi_S^s$, $\psi^{t,V}_A$, and $\psi^{t,V}_T$ are of twist-3.
For their expressions, the same form and parameters are adopted as in \cite{prd97033001}.
\section{ Form factors in the PQCD approach}\label{sec:formfactor}
Based on the $k_T$ factorization theorem,
the transition form factors can be expressed as the convolution of a
hard kernel with the distribution amplitudes of those mesons involved
in the decays in the heavy-quark and large-recoil limits.
For a review of this approach, please see Ref. \cite{ppnp5185}.
The hard kernel can be treated by PQCD at the leading order in an $\alpha_s$ expansion
(single gluon exchange as depicted in Fig. \ref{fig:semi}).
Below, we will derive the general formulas of the $B_c\rightarrow S,A,T$ transition form factors in the PQCD approach.
\subsection{$B_c\rightarrow \chi_{c0}$ form factors}\label{sec:chic0}
The $B_c\rightarrow \chi_{c0}$ form factors  are defined by \cite{prd79114018,prd73014017}
\begin{eqnarray}\label{eq:forms}
\langle S(P_2)|\bar{c}\gamma^{\mu}\gamma_5b|B_c(P_1)\rangle=[(P_1+P_2)^{\mu}-\frac{M^2-m^2}{q^2}q^{\mu}]F_+(q^2)
+\frac{M^2-m^2}{q^2}q^{\mu}F_0(q^2).
\end{eqnarray}
It is conventional to define two auxiliary form factors $f_1(q^2)$ and $f_2(q^2)$,
which are related to $F_+(q^2)$ and $F_0(q^2)$ by
\begin{eqnarray}
F_+(q^2)&=&\frac{1}{2}(f_1(q^2)+f_2(q^2)),\nonumber\\
F_0(q^2)&=&\frac{1}{2}f_1(q^2)[1+\frac{q^2}{M^2-m^2}]+\frac{1}{2}f_2(q^2)[1-\frac{q^2}{M^2-m^2}].
\end{eqnarray}
After standard calculations, we obtain their factorization formulas as follows:
\begin{eqnarray}\label{eq:f1ex}
f_1(q^2)&=&4 \sqrt{\frac{2}{3}} \pi  M^2 f_B C_f r \int_0^1dx_1dx_2\int_0^{\infty}b_1b_2db_1db_2\phi_{B_c}(x_1,b_1)\nonumber\\&&
[\psi^v_S(x_2,b_2) r(x_2-1)-\psi^s_S(x_2,b_2)(r_b-2)]\alpha_s (t_a)S_{ab}(t_a)h(\alpha_e,\beta_a,b_1,b_2)S_t(x_2)\nonumber\\&&
-[\psi^v_S(x_2,b_2)(r-2\eta x_1)+\psi^s_S(x_2,b_2)2(x_1+r_c)]\alpha_s (t_b)S_{ab}(t_b)h(\alpha_e,\beta_b,b_2,b_1)S_t(x_1),
\end{eqnarray}
\begin{eqnarray}\label{eq:f2ex}
f_2(q^2)&=&4 \sqrt{\frac{2}{3}} \pi  M^2 f_B C_f  \int_0^1dx_1dx_2\int_0^{\infty}b_1b_2db_1db_2\phi_{B_c}(x_1,b_1)\nonumber\\&&
[\psi^v_S(x_2,b_2)(2r_b-1-2r\eta (x_2-1))+\psi^s_S(x_2,b_2)2r(x_2-1)]\alpha_s (t_a)S_{ab}(t_a)h(\alpha_e,\beta_a,b_1,b_2)S_t(x_2)\nonumber\\&&
-[\psi^v_S(x_2,b_2)(-r_c+x_1)-\psi^s_S(x_2,b_2)2r]\alpha_s (t_b)S_{ab}(t_b)h(\alpha_e,\beta_b,b_2,b_1)S_t(x_1),
\end{eqnarray}
with  $r_{b,c}=\frac{m_{b,c}}{M}$.
  $\alpha_e$ and $\beta_{a,b}$ are the virtuality of the internal gluon and quark, respectively. Their expressions are
 \begin{eqnarray}
\alpha_e&=&-M^2[x_1+\eta^+r(x_2-1)][x_1+\eta^-r(x_2-1)],\nonumber\\
\beta_{a}&=&m_b^2-M^2[1+\eta^+r(x_2-1)][1+\eta^-r(x_2-1)],\nonumber\\
\beta_{b}&=&m_c^2-M^2[\eta^+r-x_1][\eta^-r-x_1],
\end{eqnarray}
where the explicit expressions of the functions $S_t$, $h$, and the scales $t_{a,b}$
 are referred to \cite{prd90114030}. The modified Sudakov factor $S_{ab}$,
  which includes the charm quark mass effect, can be found in \cite{180106145}.
\subsection{$B_c\rightarrow \chi_{c1},h_c$ form factors}\label{sec:chic1}
Following Ref. \cite{epjc76564}, the $B_c\rightarrow \chi_{c1},h_c$ transition induced by the vector and axial-vector currents is
parametrized by
\begin{eqnarray}\label{eq:formaa}
\langle A(P_2)|\bar{c}\gamma^{\mu}b|B_c(P_1)\rangle&=&2m\frac{\epsilon^*\cdot q}{q^2}q^{\mu}V_0(q^2)+
(M-m)[\epsilon^{*\mu}-\frac{\epsilon^*\cdot q}{q^2}q^{\mu}]V_1(q^2)\nonumber\\&&
-\frac{\epsilon^*\cdot q}{M-m}[(P_1+P_2)^{\mu}-\frac{M^2-m^2}{q^2}q^{\mu}]V_2(q^2),\nonumber\\
\langle A(P_2)|\bar{c}\gamma^{\mu}\gamma_5b|B_c(P_1)\rangle &=&\frac{2iA(q^2)}{M-m}\epsilon^{\mu\nu\rho\sigma}
\epsilon^*_{\nu}P_{2\rho}P_{1\sigma},
\end{eqnarray}
where the convention $\epsilon^{0123}=+1$ is taken.
Compared with the  $B_c\rightarrow J/\psi$ transition, here the
behavior of the vector and axial-vector currents  is interchanged, and the factor $M+m$ is replaced by $M-m$.
The relation  $2r V_0(0)=(1-r)A_1(0)-(1+r)A_2(0)$ is obtained  to smear the singularity at $q^2=0$.

The factorization formulas are  acquired as 
\begin{eqnarray}\label{eq:v0ex}
V_0(q^2)&=&-2 \sqrt{\frac{2}{3}} \pi  M^2 f_B C_f  \int_0^1dx_1dx_2\int_0^{\infty}b_1b_2db_1db_2\phi_{B_c}(x_1,b_1)\nonumber\\&&
[\psi ^L(x_2,b_2) \left(1-2 r_b-r(x_2-1)(r-2\eta)\right)-\psi ^t(x_2,b_2)r \left(2 x_2-r_b\right)]\alpha_s(t_a)S_{ab}(t_a)h(\alpha_e,\beta_a,b_1,b_2)S_t(x_2)\nonumber\\&&
-\psi ^L(x_2,b_2)[-r_c+r^2+x_1(1-2r\eta )]\alpha_s(t_b)S_{ab}(t_b)h(\alpha_e,\beta_b,b_2,b_1)S_t(x_1),
\end{eqnarray}
\begin{eqnarray}\label{eq:v1ex}
V_1(q^2)&=&4 \sqrt{\frac{2}{3}}\frac{r}{1-r} \pi  M^2 f_B C_f\int_0^1dx_1dx_2\int_0^{\infty}b_1b_2db_1db_2\phi_{B_c}(x_1,b_1)\nonumber\\&&
[\psi^V_A(x_2,b_2) \left(-2 r_b+\eta  r (x_2-1)+1\right)+\psi^T_A(x_2,b_2)[\eta  r_b-2 (\eta +r (x_2-1))]]\nonumber\\&&\alpha_s(t_a)S_{ab}(t_a)h(\alpha_e,\beta_a,b_1,b_2)S_t(x_2)\nonumber\\&&
-\psi^V_A(x_2,b_2)[-r_c-x_1+\eta r]\alpha_s(t_b)S_{ab}(t_b)h(\alpha_e,\beta_b,b_2,b_1)S_t(x_1),
\end{eqnarray}
\begin{eqnarray}\label{eq:v2ex}
V_2(q^2)&=&-A_1\frac{(1-r)^2(r-\eta)}{2r(\eta^2-1)}-2 \pi  M^2 f_B C_f\sqrt{\frac{2}{3}}\frac{1-r}{\eta^2-1}
\int_0^1dx_1dx_2\int_0^{\infty}b_1b_2db_1db_2\phi_{B_c}(x_1,b_1)\nonumber\\&&
[\psi^t_A(x_2,b_2)(r_b (1-\eta  r)+2 r^2 (x_2-1)-2 \eta  r (x_2-2)-2)\nonumber\\&&
-\psi^L_A(x_2,b_2)(2 r_b (\eta -r)-\eta +r (\eta  r (x_2-1)-2 \eta ^2 (x_2-1)+x_2))]
\nonumber\\&&\alpha_s(t_a)S_{ab}(t_a)h(\alpha_e,\beta_a,b_1,b_2)S_t(x_2)\nonumber\\&&
+\psi_A^L(x_2,b_2)[-r_c (r-\eta )+\eta  r^2+r \left(-2 \eta ^2 x_1+x_1-1\right)+\eta  x_1]
\nonumber\\&&\alpha_s(t_b)S_{ab}(t_b)h(\alpha_e,\beta_b,b_2,b_1)S_t(x_1),
\end{eqnarray}
\begin{eqnarray}\label{eq:Aex}
A(q^2)&=&2\sqrt{\frac{2}{3}}\pi  M^2 f_B C_f (1-r)\int_0^1dx_1dx_2\int_0^{\infty}b_1b_2db_1db_2\phi_{B_c}(x_1,b_1)\nonumber\\&&
[\psi^V_A(x_2,b_2) r\left(1-x_2\right)+\psi^T_A(x_2,b_2)(r_b-2)]\alpha_s(t_a)S_{ab}(t_a)h(\alpha_e,\beta_a,b_1,b_2)S_t(x_2)\nonumber\\&&
-\psi^V_A(x_2,b_2)r\alpha_s(t_b)S_{ab}(t_b)h(\alpha_e,\beta_b,b_2,b_1)S_t(x_1).
\end{eqnarray}
It should be stressed that    the nonlocal matrix element for the axial-vector and  scalar  charmonium meson in Eq. (\ref{eq:non})
can be  related to the vector and pseudoscalar ones \cite{epjc76564}, respectively,
 by multiplying by the structure $(-i)\gamma_5$ from the left-hand side.
The factorization formulas  $f_{1,2}$, $V_{0,1,2}$, and $A$ here are similar to
the corresponding ones in \cite{epjc76564} with the $r_c$ term flipping signs  
and  the replacement $1+r\rightarrow1-r$.
\subsection{$B_c\rightarrow \chi_{c2}$ form factors}\label{sec:chic2}
In analogy with $B_c\rightarrow J/\psi$ form factors, we parametrize the $B_c\rightarrow \chi_{c2}$ form factors induced by the vector and axial-vector currents as
\begin{eqnarray}\label{eq:form2}
\langle T(P_2)|\bar{c}\gamma^{\mu}b|B_c(P_1)\rangle&=&\frac{2iV(q^2)}{M+m}\epsilon^{\mu\nu\rho\sigma}
\epsilon^*_{T\nu}P_{2\rho}P_{1\sigma},\nonumber\\
\langle T(P_2)|\bar{c}\gamma^{\mu}\gamma_5b|B_c(P_1)\rangle&=&2m\frac{\epsilon^*_T\cdot q}{q^2}q^{\mu}A_0(q^2)+
(M+m)[\epsilon^{*\mu}_T-\frac{\epsilon^*_T\cdot q}{q^2}q^{\mu}]A_1(q^2)\nonumber\\&&
-\frac{\epsilon^*_T\cdot q}{M+m}[(P_1+P_2)^{\mu}-\frac{M^2-m^2}{q^2}q^{\mu}]A_2(q^2).
\end{eqnarray}
Note that the structure of above form factors is analogous to the $J/\psi$ case with the replacement $\epsilon \rightarrow\epsilon_T$.
 In addition, as mentioned before, the LCDAs of a tensor meson are also similar to the vector ones except that the  $\epsilon$ is replaced  by $\epsilon_{\bullet}$.
So, the factorization formulas here  can be straightforwardly obtained  by replacing the twist-2 or
twist-3 LCDAs of the $J/\psi$ with the corresponding twists of the $\chi_{c2}$ one in Eq. (\ref{eq:non}).
After multiplying by the different definitions of the polarization vector, we have \cite{prd83014008}
 \begin{eqnarray}\label{eq:form3}
\mathcal{F}^{B_c\rightarrow \chi_{c2}}=\frac{\epsilon_{\bullet}}{\epsilon_T}
\mathcal{F}^{B_c\rightarrow J/\psi}|_{\psi_V\rightarrow \psi_T}=\frac{1}{\sqrt{\eta^2-1}}
\mathcal{F}^{B_c\rightarrow J/\psi}|_{\psi_V\rightarrow \psi_T}.
\end{eqnarray}
\subsection{The semileptonic differential decay rates}\label{sec:decay}
As is well known, the above form factors are reliable only in the small $q^2$ region in the PQCD framework \cite{prd83014008,prd79034014}.
In order to estimate the  semileptonic differential decay rates, we need to know the $q^2$-dependent
form factors in the full  kinematical  region. Our form factors are truncated at about $q^2=m^2_{\tau}$ with $m_{\tau}$ the mass of the $\tau$ lepton. We first perform the PQCD calculations on them in the  range of $0<q^2<m^2_{\tau}$, while
the momentum dependence of the  form factors in the $m^2_{\tau}<q^2<(M-m)^2$ region 
 is determined by  fitting 
 through a three-parameter function. 
 The following fit parametrization is chosen for the form factors with respect to $q^2$ \cite{epjc76564}:
 \begin{eqnarray}\label{eq:extr}
\mathcal {F}_i(q^2)=\mathcal {F}_i(0)\exp[a\frac{q^2}{M^2}+b(\frac{q^2}{M^2})^2],
\end{eqnarray}
where $\mathcal {F}_i$ denotes any one of the form factors, and $a$, $b$ are the fitted  parameters.

After integrating out the off-shell $W$ boson,
the effective Hamiltonian for the $b\rightarrow cl \nu_l$ transition is written as \cite{rmp1125}
\begin{eqnarray}
\mathcal{H}_{eff}=\frac{G_F}{\sqrt{2}}V^*_{cb}\bar{b}\gamma_{\mu}
(1-\gamma_5)c\otimes \bar{\nu_l}\gamma^{\mu}(1-\gamma_5)l,
\end{eqnarray}
where $G_F=1.16637 \times 10^{-5} \text{GeV}^{-2} $ is the Fermi coupling
constant and $V_{cb}$ is one of the CKM matrix elements.
The differential decay rate of the exclusive processes $B_c\rightarrow (S,A) l \nu$
can be expressed in terms of the form factors as \cite{prd79114018}
\begin{eqnarray}\label{eq:ds}
\frac{d\Gamma}{dq^2}(B_c\rightarrow S l \nu)=\frac{G_F^2|V_{cb}|^2}{384\pi^3M^3q^2}\sqrt{\lambda(q^2)}
(1-\frac{m_l^2}{q^2})^2[3m_l^2(M^2-m^2)^2|F_0(q^2)|^2
+(m_l^2+2q^2)\lambda(q^2)|F_+(q^2)|^2],
\end{eqnarray}
\begin{eqnarray}\label{eq:da2}
\frac{d\Gamma_{L}}{dq^2}(B_c\rightarrow A l \nu)&=&\frac{G_F^2|V_{cb}|^2}{384\pi^3M^3q^2}\sqrt{\lambda(q^2)}
(1-\frac{m_l^2}{q^2})^2\nonumber\\&&
\{3m_l^2\lambda(q^2)|V_0(q^2)|^2
+\frac{m_l^2+2q^2}{4m^2}|(M^2-m^2-q^2)(M-m)V_1(q^2)-
\frac{\lambda(q^2)}{M-m}V_2(q^2)|^2\},
\end{eqnarray}
\begin{eqnarray}\label{eq:da1}
\frac{d\Gamma_{\pm}}{dq^2}(B_c\rightarrow A l \nu)&=&\frac{G_F^2|V_{cb}|^2}{384\pi^3M^3}\lambda^{3/2}(q^2)
(1-\frac{m_l^2}{q^2})^2(m_l^2+2q^2)|\frac{A(q^2)}{M-m}
\mp\frac{(M-m)V_1(q^2)}{\sqrt{\lambda(q^2)}}|^2,
\end{eqnarray}
where $m_l$ is the lepton mass  and $\lambda(q^2)=(M^2+m^2-q^2)^2-4M^2m^2$.
The subscripts $L$, $+$, and $-$ 
denote the longitudinal, positive, and negative polarizations of the final state, respectively.
As stated before,  the decay width of  $B_c\rightarrow \chi_{c2} l\nu$ can be related to the $J/\psi$ one \cite{epjc76564} by making the following replacement:
\begin{eqnarray}\label{eq:dt1}
\frac{d\Gamma_{L}}{dq^2}(B_c\rightarrow \chi_{c2} l \nu)&=& \frac{2(\eta^2-1)}{3}
\frac{d\Gamma_{L}}{dq^2}(B_c\rightarrow J/\psi l \nu)|
_{\mathcal{F}^{B_c\rightarrow J/\psi}\rightarrow \mathcal{F}^{B_c\rightarrow \chi_{c2}}},\nonumber\\
\frac{d\Gamma_{\pm}}{dq^2}(B_c\rightarrow \chi_{c2} l \nu)&=& \frac{\eta^2-1}{2}
\frac{d\Gamma_{\pm}}{dq^2}(B_c\rightarrow J/\psi l \nu)|
_{\mathcal{F}^{B_c\rightarrow J/\psi}\rightarrow \mathcal{F}^{B_c\rightarrow \chi_{c2}}},
\end{eqnarray}
where the factors  $\frac{2(\eta^2-1)}{3}$ and $\frac{\eta^2-1}{2}$ come from Eq.(\ref{eq:vvv}).
The total differential widths for the axial-vector and tensor charmonium modes can be written as
\begin{eqnarray}\label{eq:all}
 \frac{d\Gamma}{dq^2}= \frac{d\Gamma_{L}}{dq^2}+\frac{d\Gamma_{+}}{dq^2}+ \frac{d\Gamma_{-}}{dq^2}.
\end{eqnarray}
\section{ Numerical analysis and discussion}\label{sec:results}
For numerical evaluation, we collect the input parameters such as the  masses  and  the
meson decay constants in Table \ref{tab:constant1}, 
while the CKM matrix elements and $B_c$ lifetime  are set as $V_{cb}=0.0405$ \cite{pdg2016}
and $\tau_{B_c}=0.507$ ps \cite{pdg2016}, respectively.
In the fitting procedure, the form factors in the lower region, namely, $q^2\in [0,m_{\tau}^2]$,
are computed in the  PQCD framework.
The numerical results of the relevant form factors at the scale $q^2=0$ as well as the
fitted parameters $a$ and $b$ are presented in Table \ref{tab:form},
 and here the uncertainties for our results are estimated including three aspects.
 The first type of error comes from the shape parameter $\omega$ of the $B_c$ meson distribution amplitude;
 the second one is from the charm quark mass; the last one is caused by the decay constants of the charmonium states.
 In the evaluation, 
these uncertainties are obtained by simply taking a $\pm10\%$  uncertainty on the central value.
The combined uncertainties can reach   $25\%$. In
addition, the uncertainties from the CKM matrix elements and the hard scale $t$ are very small and have been neglected.
\begin{table}
\caption{The quark masses and the $B_c$ meson decay constant are  taken from \cite{180106145},  while
 the decay constants of the  $P$-wave charmonium states are adopted from the   recent updated values evaluated from the QCD
  sum rules at the scales $\mu=m_c$  \cite{prd96014026}.
Other parameters are from PDG 2016 \cite{pdg2016}. }
\label{tab:constant1}
\begin{tabular*}{18cm}{@{\extracolsep{\fill}}llccc}
  \hline\hline
\text{Mass(\text{GeV})}
& $M=6.277$  & $m_{\tau}=1.777$  &$m_{b}=4.8$ & $m_{c}=1.5$ \\[1ex]
& $m_{\chi_{c0}}=3.415$ & $m_{\chi_{c1}}=3.3.511$  & $m_{\chi_{c2}}=3.556$ & $m_{h_c}=3.525$ \\[1ex]
\end{tabular*}
\begin{tabular*}{18cm}{@{\extracolsep{\fill}}lcccc}
\text{Decay constants (MeV)}
& $f_{B_c}=0.489$ & $f_{\chi_{c0}}=0.0916$  & $f_{\chi_{c1}}=0.185$  & $f_{\chi_{c1}}^{\perp}=0.0875$   \\[1ex]
& $f_{\chi_{c2}}=0.177$  & $f_{\chi_{c2}}^{\perp}=0.128$  & $f_{h_c}=0.127$  & $f_{h_c}^{\perp}=0.133$ \\[1ex]
\hline\hline
\end{tabular*}
\end{table}

\begin{table}
\caption{The form factors of  the $B_c$ meson decay to $P$-wave charmonium  evaluated by PQCD and by other methods in the literature. We
also show theoretical uncertainties induced by the shape parameter $\omega$, charm quark mass $m_c$, and the decay constants of charmonium states,  respectively.
The last two columns correspond to the fit parameters $a$ and $b$ in this work.}
\label{tab:form}
\begin{tabular}[t]{lccccccc}
\hline\hline
$\mathcal{F}_i$    &This work  & QCDSR \cite{prd79116001}    &LFQM \cite{prd79114018}   &NRQCD \cite{171007011}
\footnotemark[1]   &ISGW II \cite{prd87034004}     & $a$ & $b$ \\ \hline
$F_{0}^{B_c\rightarrow \chi_{c0}}$ & $0.41^{+0.09+0.01+0.04}_{-0.07-0.02-0.04}$   &$0.673\pm0.195$   &$0.47^{+0.03}_{-0.06}$    &$1.25^{+0.15}_{-0.12}$  &$\cdots$       & 2.6&2.8  \\
$F_{+}^{B_c\rightarrow \chi_{c0}}$ & $0.41^{+0.09+0.01+0.04}_{-0.07-0.02-0.04}$     &$0.673\pm0.195$   &$0.47^{+0.03}_{-0.06}$    &$1.25^{+0.15}_{-0.12}$  &$\cdots$         & 3.6&2.6  \\\hline
$A^{B_c\rightarrow \chi_{c1}}$     & $0.18^{+0.03+0.01+0.02}_{-0.03-0.02-0.02}$   &$0.13\pm0.04$     &$0.36^{+0.02}_{-0.04}$    &$0.99^{+0.19}_{-0.15}$  &$-0.36^{+0.01}_{-0.01}$      & 2.4&13.8  \\
$V_0^{B_c\rightarrow \chi_{c1}}$   & $0.18^{+0.02+0.01+0.02}_{-0.02-0.02-0.02}$   &$0.03\pm0.01$     &$0.13^{+0.01}_{-0.01}$     &$0.12^{+0.01}_{-0.01}$ &$-0.55^{+0.00}_{-0.01}$        & 4.8&$-0.2$  \\
$V_1^{B_c\rightarrow \chi_{c1}}$    & $0.86^{+0.14+0.06+0.09}_{-0.10-0.08-0.09}$  &$0.30\pm0.09$     &$0.85^{+0.02}_{-0.04}$    &$2.34^{+0.21}_{-0.22}$  &$-0.42^{+0.02}_{-0.02}$                  & 2.7&$-11.0$  \\
$V_2^{B_c\rightarrow \chi_{c1}}$   & $0.11^{+0.02+0.01+0.01}_{-0.01-0.00-0.01}$   &$0.06\pm0.02$     &$0.15^{+0.01}_{-0.01}$      &$0.47^{+0.07}_{-0.06}$ &$0.28^{+0.01}_{-0.01}$          & 5.3&$-1.8$  \\\hline
$V^{B_c\rightarrow \chi_{c2}}$ & $1.15^{+0.15+0.00+0.11}_{-0.13-0.03-0.10}$        &$\cdots$            &$1.36^{+0.12}_{-0.19}$     &$5.89^{+1.60}_{-1.30}$  &$\cdots$          & 5.1&12.9  \\
$A_0^{B_c\rightarrow \chi_{c2}}$ & $0.83^{+0.13+0.04+0.08}_{-0.11-0.05-0.08}$      &$\cdots$             &$0.86^{+0.14}_{-0.13}$     &$1.80^{+0.40}_{-0.33}$  &$\cdots$           & 7.0&15.3  \\
$A_1^{B_c\rightarrow \chi_{c2}}$ & $0.55^{+0.08+0.01+0.06}_{-0.06-0.02-0.06}$      &$\cdots$             &$0.81^{+0.10}_{-0.10}$     &$1.95^{+0.43}_{-0.35}$   &$\cdots$       & 4.3&7.3 \\
$A_2^{B_c\rightarrow \chi_{c2}}$ & $-0.14^{+0.06+0.07+0.01}_{-0.09-0.08-0.01}$     &$\cdots$            &$0.68^{+0.06}_{-0.00}$     &$2.24^{+0.51}_{-0.42}$   &$\cdots$        & 37.1&$-96.1$  \\\hline
$A^{B_c\rightarrow h_c}$        & $0.10^{+0.02+0.00+0.01}_{-0.01-0.00-0.01}$       &$0.13\pm0.04$     &$0.07^{+0.01}_{-0.01}$    &$0.07^{+0.00}_{-0.01}$  &$0.05^{+0.00}_{-0.00}$              & 3.0&$-0.2$  \\
$V_0^{B_c\rightarrow h_c}$      & $0.22^{+0.03+0.02+0.03}_{-0.02-0.01-0.02}$       &$0.03\pm0.01$     &$0.64^{+0.10}_{-0.02}$   &$1.63^{+0.25}_{-0.19}$  &$0.78^{+0.01}_{-0.01}$                  & 3.1&1.8  \\
$V_1^{B_c\rightarrow h_c}$      & $0.46^{+0.05+0.01+0.05}_{-0.05-0.03-0.05}$       &$0.30\pm0.09$     &$0.50^{+0.05}_{-0.08}$   &$0.46^{+0.07}_{-0.03}$  &$0.61^{+0.01}_{-0.01}$               & 2.6&$-1.1$  \\
$V_2^{B_c\rightarrow h_c}$      & $-0.03^{+0.00+0.00+0.00}_{-0.00-0.01-0.00}$      &$0.06\pm0.02$     &$-0.32^{+0.06}_{-0.05}$  &$-0.75^{+0.17}_{-0.17}$  &$-0.39^{+0.01}_{-0.01}$                & 7.5&41.1  \\
\hline\hline
\end{tabular}
\footnotetext[1]{We quote the leading-order  results of NRQCD.}
\end{table}

It is found that the form factors of the $P$-wave modes 
are smaller than those of the $S$-wave ones  in our previous study \cite{epjc76564}.
This phenomenon can be understood from the wave functions of the two states.
The additional nodes in the wave functions of the orbital excited charmonium state  cause the overlap between the initial
and final state wave functions to become smaller. In addition,
 the smaller decay constants of $P$-wave charmonium states  also suppress the corresponding values.
Comparing the form factors of $B_c\rightarrow \chi_{c1}$ with $B_c\rightarrow h_c$ in Table. \ref{tab:form},
one can find the large differences  between them.
The main reason is  the different  DAs and the decay constants for the two kinds of axial-vector charmonium.
Because of  the G parity, the DAs for $\chi_{c1}$ and $h_c$ mesons exhibit the different asymptotic behaviors \cite{prd97033001}.
Moreover,  the longitudinal and transverse decay constants (see Table \ref{tab:constant1})
in the two axial-vector mesons can also contribute to different values.
The $B_c\rightarrow T$ transition  form  factor  is  somewhat larger since the prefactor in Eq.(\ref{eq:form3})
is roughly $2r/(1-r^2)\approx 1.7$ at the maximally recoiling point, which enhanced the numbers accordingly.

So far, several authors have calculated the form factors of
the concerned decays via different frameworks.
To compare the results, we should rescale them
according to the form factor definitions in Eqs. (\ref{eq:forms}), (\ref{eq:formaa}), and (\ref{eq:form2}).
For example, comparing the definitions of the $B_c\rightarrow T$ transition form factor of Ref. \cite{prd79114018} with ours,
we have the following relations at the maximal recoil point:
 \begin{eqnarray}
 V=M(M+m)h,\quad A_1=\frac{M}{M+m}k,\quad A_2=-M(M+m)b_+,
\end{eqnarray}
where the values of $h$, $k$, and $b_+$ can be found in \cite{prd79114018}.
Note that  we have dropped an overall phase factor $i$
which is irrelevant for the calculation of the decay widths.
Other results, such as QCDSR \cite{prd79116001}, ISGW II \cite{prd87034004}, and NRQCD \cite{171007011},
 are also converted into the  numbers according to our definitions in this paper  and are listed in Table \ref{tab:form}.
As indicated in Table \ref{tab:form}, the results evaluated in the different models are roughly comparable.
Our results are generally close to those of the LFQM \cite{prd79114018} and the QCDSR \cite{prd79116001},
while some of the results for the $B_c\rightarrow \chi_{c1}$ transition from the ISGW II model
 possess a sign that is  the opposite of ours.
The recent NRQCD predictions in \cite{171007011} are obviously larger for most of the decay channels.

\begin{figure}[tbp]
\begin{center}
\setlength{\abovecaptionskip}{0pt}
\centerline{
\hspace{4cm}\subfigure{\epsfxsize=15 cm \epsffile{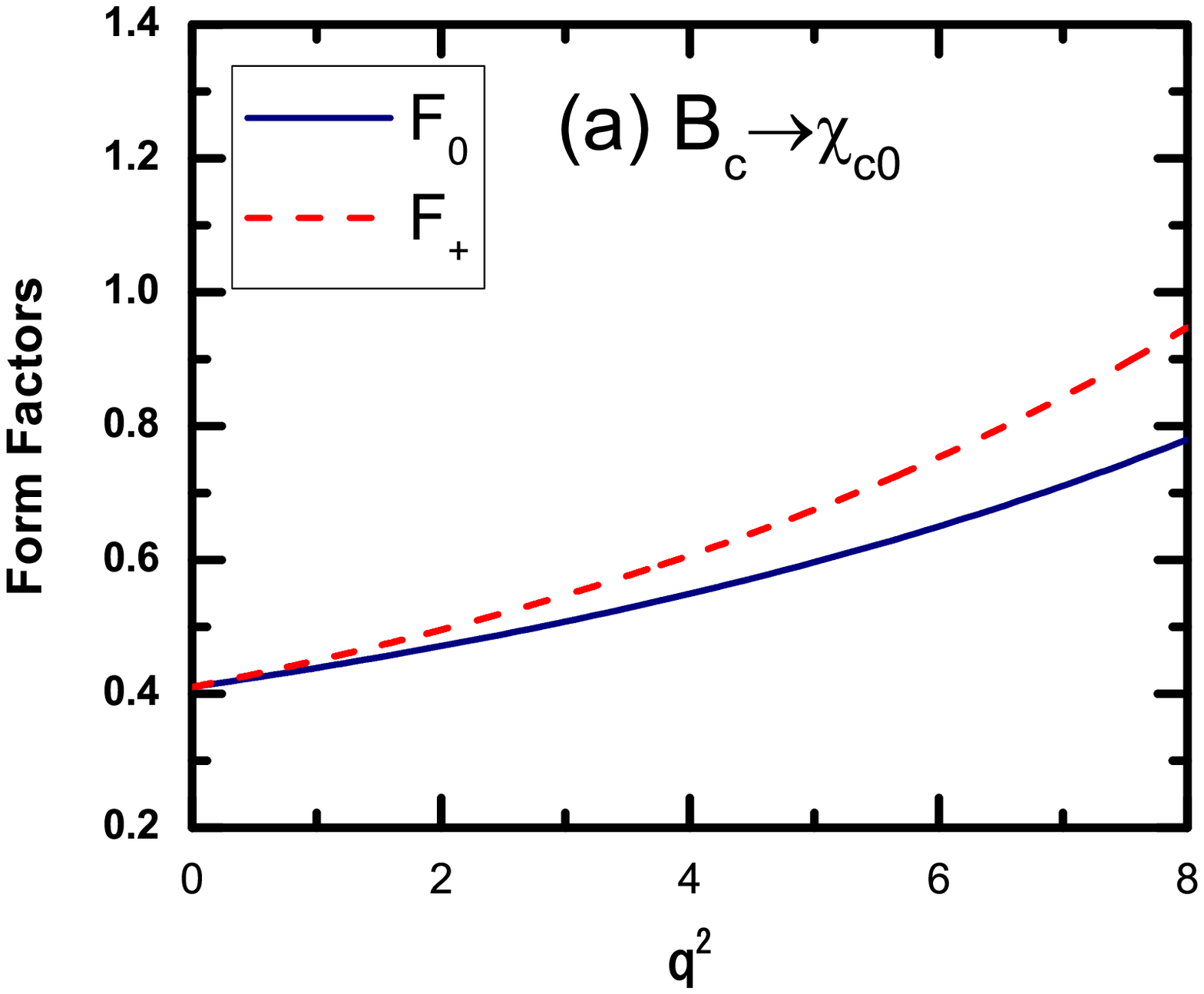} }
\hspace{-6cm}\subfigure{ \epsfxsize=15 cm \epsffile{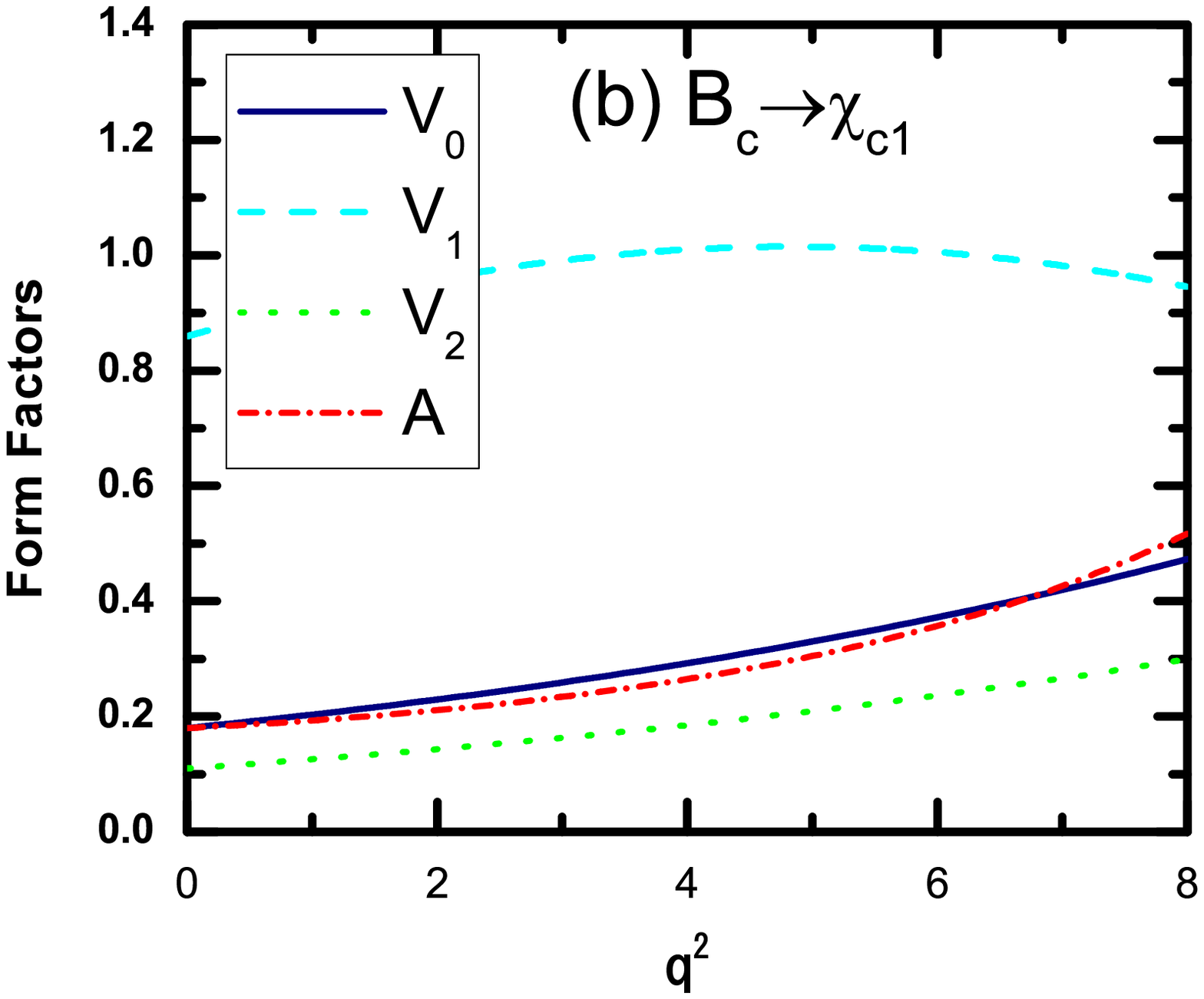}}}
\vspace{-3cm}
\centerline{
\hspace{4cm}\subfigure{\epsfxsize=15 cm \epsffile{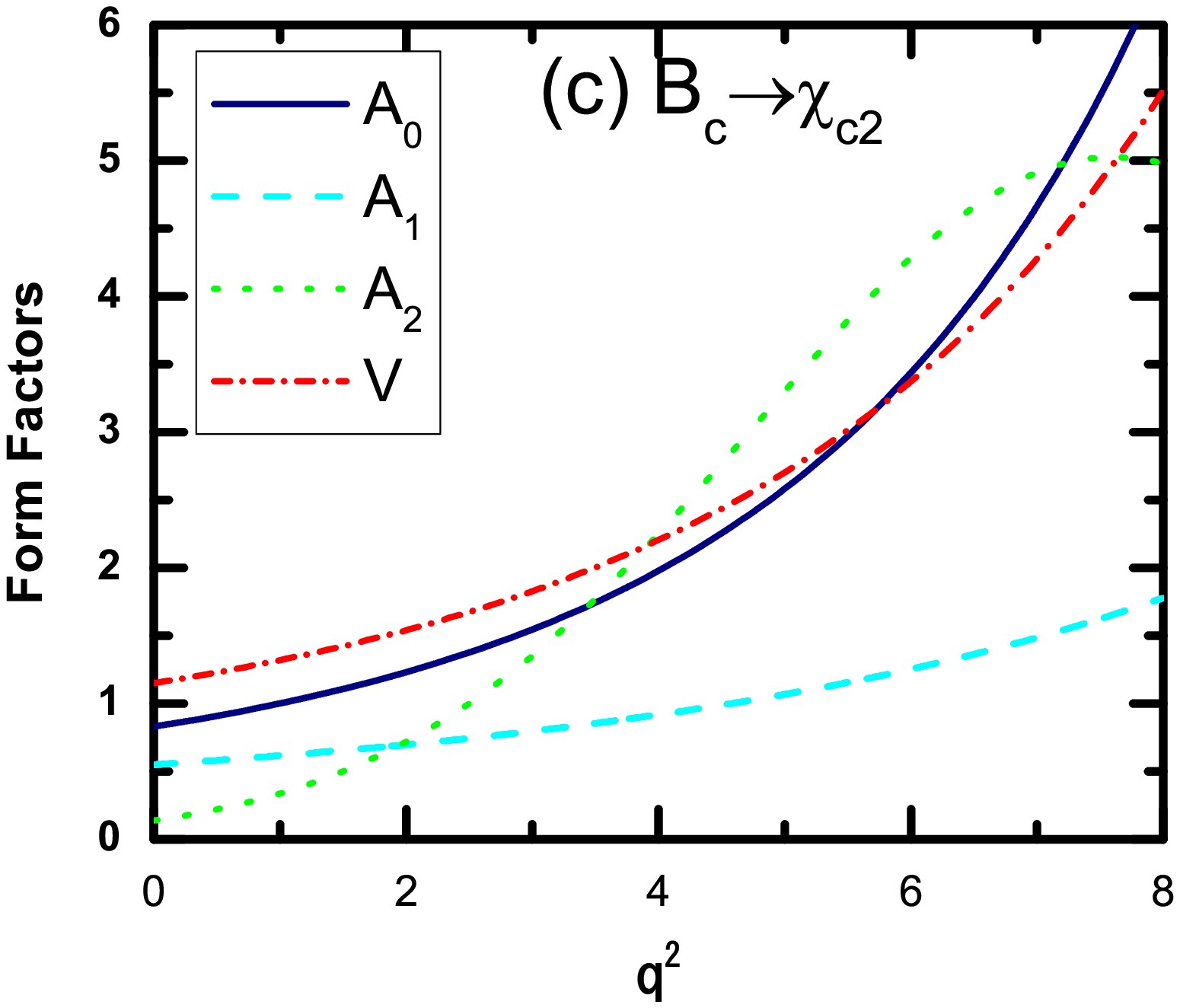} }
\hspace{-6cm}\subfigure{ \epsfxsize=15 cm \epsffile{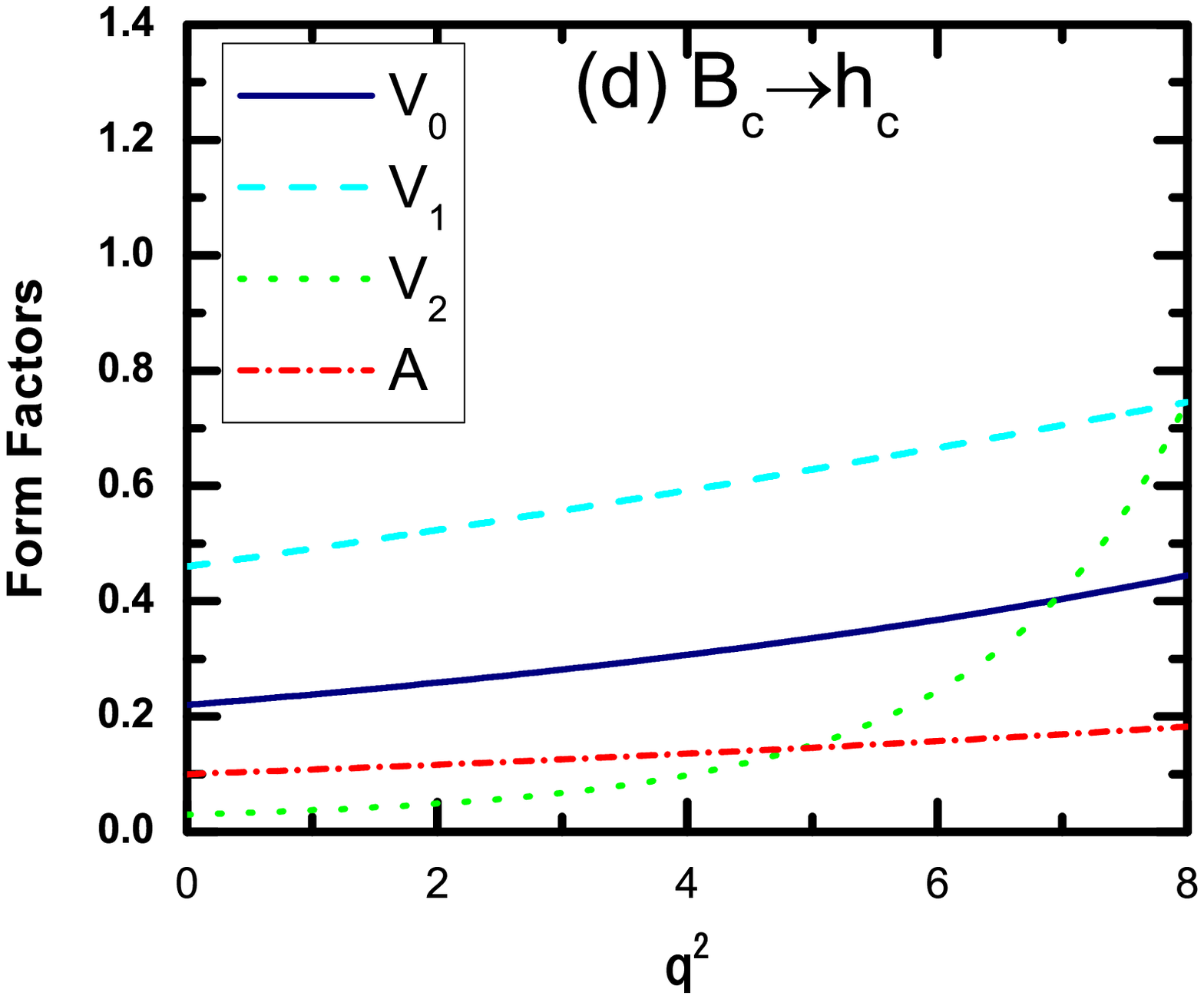}}}
\vspace{-2cm}\caption{The $q^2$ dependence of the transition form factors for the decay modes (a) $B_c\rightarrow \chi_{c0}$, (b) $B_c\rightarrow \chi_{c1}$,   (c) $B_c\rightarrow \chi_{c2}$, and (d) $B_c\rightarrow h_c$.
A minus sign has been added to $A_2^{B_c\rightarrow\chi_{c2}}$ and $V_2^{B_c\rightarrow h_{c}}$
so that the corresponding curves show in the upper panels.}
 \label{fig:reda}
\end{center}
\end{figure}

Based on the values of the  transition form factors at $q^2=0$ and the fit parameters $a$ and $b$ listed in Table \ref{tab:form},
 we can plot the momentum transfer squared dependence of  these form factors  in Fig. \ref{fig:reda} for the four processes
 in the whole accessible kinematical range. 
 The difference of the curve behavior for the various $P$-wave charmonium states is the
consequence of their different LCDAs.
 The form factors for the $B_c\rightarrow \chi_{c2}$ transition have a relatively stronger  momentum dependence  than others.
The main reason is that the $B_c\rightarrow \chi_{c2}$ form factors received additional $q^2$ dependence as can be seen from
the factorization formulas in Eq.(\ref{eq:form3}), which provide an enhancement to  the corresponding values with the increase of  $q^2$.

With the form factors at hand, one can directly obtain the partial decay width
 by integrating the corresponding differential decay rates over $q^2$ in Eqs. (\ref{eq:ds})-(\ref{eq:dt1}).
We are now ready to calculate the respective semileptonic decay branching ratios.
The numerical results are shown in Table \ref{tab:br},
together with the numbers obtained in other model calculations for comparison.
In general, it is observed that  the branching ratios have close values within the error bars in all models.
In particular, our results match very well with those of QCDSR \cite{prd79116001}.

Since the electron and muon are very light compared with the heavy tau lepton,
we  neglect  their masses in the calculations.
It is seen that the semitauonic decays branching ratios fall short by a large factor
compared with
the corresponding values of the  $e$ and $\mu$ channels due to suppression from the phase space.
In order to reduce the theoretical uncertainties from the hadronic parameters,
 we define four
 ratios between the  branching fractions of semitauonic decays of $B_c$ mesons relative to the
decays involving lighter lepton families,
\begin{eqnarray}\label{eq:ratio1}
\mathcal {R}(X)=\frac{\mathcal {B}(B_c^+\rightarrow X \tau^+\nu_{\tau})}{\mathcal {B}(B_c^+\rightarrow X e^+ \nu_e)}.
\end{eqnarray}
From the numbers in Table \ref{tab:br}, we have
  \begin{eqnarray}
\mathcal {R}(\chi_{c0}) =0.22^{+0.00}_{-0.01},\quad
\mathcal {R}(\chi_{c1}) =0.13^{+0.01}_{-0.00},\quad
\mathcal {R}(\chi_{c2}) =0.08^{+0.01}_{-0.00},\quad
\mathcal {R}(h_c) =0.12^{+0.01}_{-0.00},
\end{eqnarray}
where all uncertainties are added in quadrature. 
The central values  lie between 0.08 and 0.22, 
which are typically smaller than our previous prediction for that of $J/\psi$ with $\mathcal {R}(J/\psi)=0.29$ \cite{epjc76564}
because the heavy $P$-wave charmonium states  bring a smaller phase space than the $S$-wave ones.
More recently, the LHCb Collaboration \cite{prl120121801}  published a measurement $\mathcal {R}(J/\psi)=0.71$ that
shows the discrepancy with the prediction of the SM.
It would be interesting to see whether the similar anomalies  also exist independently in these $P$-wave charmonium modes.
Therefore, the measurements of various  ratios such as $\mathcal{R}(X)$
in the future will give an additional hint for the NP effect in the $b\rightarrow c l \nu$ transition.
\begin{table}\footnotesize
\caption{Branching ratios (in units of $10^{-3}$) of semileptonic  $B_c$ decays evaluated by PQCD and by other methods in the literature.
The errors are induced by the same sources as in Table \ref{tab:form}. }
\label{tab:br}
\begin{tabular}[t]{lccccccccc}
\hline\hline
Modes   & This Work  & QCDSR \cite{prd79116001}    &LFQM \cite{prd79114018}  &RGM \cite{prd65014017}
&RCQM \cite{prd73054024}  &RQM \cite{prd71094006} &NRQM \cite{prd74074008} &BS \cite{jpg39015009} &RQMQP \cite{prd82034019} \\ \hline
$B_c\rightarrow \chi_{c0} e \nu_e$ & $2.22^{+1.08+0.11+0.47}_{-0.69-0.21-0.42}$   &$1.82\pm0.51$  &$2.1^{+0.2}_{-0.4}$  & 1.2 &1.7 &1.8 &1.1&$1.3\pm0.3$   &0.87\\
$B_c\rightarrow \chi_{c0} \tau \nu_{\tau}$ & $0.48^{+0.23+0.02+0.10}_{-0.15-0.05-0.09}$ &$0.49\pm0.16$ &$0.24^{+0.01}_{-0.03}$ & 0.17& 0.13&0.18&0.13&$0.16\pm0.08$ &0.075\\
$B_c\rightarrow \chi_{c1} e \nu_e$ & $1.53^{+0.57+0.24+0.32}_{-0.35-0.30-0.29}$ &$1.46\pm0.42$&$1.4^{+0.0}_{-0.1}$ & 1.5&0.92&0.98&0.66&$1.1\pm0.3$ &0.82\\
$B_c\rightarrow \chi_{c1} \tau \nu_{\tau}$ & $0.20^{+0.08+0.03+0.04}_{-0.04-0.03-0.04}$ &$0.147\pm0.044$ &$0.15^{+0.01}_{-0.02}$ &0.24&0.089&0.12&0.072&$0.097\pm0.065$&0.092\\
$B_c\rightarrow \chi_{c2} e \nu_e$ & $2.68^{+1.23+0.50+0.56}_{-0.80-0.57-0.51}$ &... &$1.7^{+0.5}_{-0.7}$ &1.9&1.7&2.0&1.3&$1.0\pm0.3$&1.6\\
$B_c\rightarrow \chi_{c2} \tau \nu_{\tau}$ & $0.22^{+0.09+0.03+0.04}_{-0.06-0.04-0.04}$&...&$0.096^{+0.027}_{-0.036}$ &0.29&0.082&0.14&0.093&$0.082\pm0.048$&0.093\\
$B_c\rightarrow h_c e \nu_e$ & $1.06^{+0.19+0.12+0.22}_{-0.21-0.13-0.20}$  &$1.42\pm0.40$&$3.1^{+0.5}_{-0.8}$ &1.8&2.7&3.1&1.7&$2.8\pm0.8$&0.96\\
$B_c\rightarrow h_c \tau \nu_{\tau}$ & $0.13^{+0.02+0.01+0.03}_{-0.03-0.02-0.02}$ &$0.137\pm0.038$&$0.22^{+0.02}_{-0.04}$&0.25 &0.17&0.27&0.15&$0.19\pm0.13$&0.077\\
\hline\hline
\end{tabular}
\end{table}

 \begin{table}
\caption{The PQCD predictions for the  polarization fractions. The errors are induced by the same sources as in Table \ref{tab:form}.}
\label{tab:polar}
\begin{tabular}[t]{lccc}
\hline\hline
Modes     &$f_{0}$ &$f_{+}$&$f_-$  \\ \hline
$B_c \rightarrow \chi_{c1}e\nu_e$  & $0.34^{+0.01+0.01+0.00}_{-0.00-0.01-0.00}$&$0.04^{+0.01+0.00+0.00}_{-0.00-0.00-0.00}$
&$0.62^{+0.00+0.00+0.00}_{-0.02-0.01-0.00}$      \\
$B_c \rightarrow \chi_{c1}\tau\nu_{\tau}$  & $0.33^{+0.00+0.00+0.00}_{-0.01-0.01-0.00}$&$0.06^{+0.01+0.00+0.00}_{-0.00-0.00-0.00}$
&$0.61^{+0.01+0.01+0.00}_{-0.01-0.00-0.00}$      \\
$B_c \rightarrow \chi_{c2}e\nu_e$  & $0.77^{+0.03+0.04+0.00}_{-0.02-0.03-0.00}$&$0.03^{+0.01+0.01+0.00}_{-0.00-0.00-0.00}$
&$0.20^{+0.03+0.04+0.00}_{-0.02-0.02-0.00}$      \\
$B_c \rightarrow \chi_{c2}\tau\nu_{\tau}$  & $0.72^{+0.02+0.03+0.00}_{-0.03-0.04-0.00}$&$0.05^{+0.01+0.01+0.00}_{-0.00-0.00-0.00}$
&$0.23^{+0.02+0.02+0.00}_{-0.02-0.03-0.00}$      \\
$B_c \rightarrow h_{c}e\nu_e$      & $0.68^{+0.01+0.03+0.00}_{-0.02-0.00-0.00}$&$0.04^{+0.00+0.00+0.00}_{-0.01-0.00-0.00}$
&$0.28^{+0.02+0.01+0.00}_{-0.01-0.02-0.00}$      \\
$B_c \rightarrow h_{c}\tau\nu_{\tau}$      & $0.60^{+0.00+0.02+0.00}_{-0.02-0.00-0.00}$&$0.07^{+0.00+0.00+0.00}_{-0.00-0.00-0.00}$
&$0.33^{+0.02+0.00+0.00}_{-0.00-0.02-0.00}$      \\
\hline\hline
\end{tabular}
\end{table}

Next, 
we made a comprehensive polarization analysis of the axial-vector and tensor channels.
Since the initial state $B_c$ is a spinless particle,
the final state axial-vector/tensor charmonium and lepton pair  carry spin degrees of freedom.
 According to the angular momentum conservation, the semileptonic decays of $B_c\rightarrow A/T l\nu_l$
 contain three different polarizations. 
 It is meaningful to define three  polarization fractions $f_{L,\pm}=\Gamma_{L,\pm}/(\Gamma_L+\Gamma_++\Gamma_-)$.
 Their individual  polarization fractions  are shown in Table \ref{tab:polar},
 where the sources of the errors in the numerical estimates
have the same origin as in Table \ref{tab:form}.
We made the following observations. First,
the minus polarization fractions have larger magnitudes in comparison
to the plus components, and the latter are only 
at the percent level.
From  Table \ref{tab:form}, one can see the
form factors $A$ and $V_1$ have the same sign, which
gives constructive contributions to the minus polarized decay width
 but destructive contributions to the plus partners as can be seen in Eq. (\ref{eq:da1}).
 Second, for $B_c\rightarrow \chi_{c1}$ decays,  the transverse
polarization contributions dominated the branching ratio due to
a destructive interference  between $V_1$ and $V_2$ in the longitudinally polarized decay width. However,
in the case of the  $B_c\rightarrow h_c$ transition, the value of $V_2$ is a negative number,
which reverses the constructive or destructive interference situation.
The dramatically different polarization contributions between the two axial-vector decay channels are similar to
the explanation in \cite{prd79114018}. Finally,  for each charmonium channel,
the longitudinal, plus, and minus polarization fractions of the $\tau$ are  roughly equal to the corresponding values of $e$,
which reflects that  the relative polarization contributions still favor the  lepton flavor universality.
These results will be tested at the ongoing and forthcoming hadron colliders.


\section{ conclusion}\label{sec:sum}

Semileptonic charmonium  decays of $B_c$ mesons play a critical role in the determination of the
magnitudes of the CKM matrix elements $V_{cb}$, and in the test of the lepton flavor universality which is
a basic assumption of the SM.
The investigation of the corresponding $P$-wave charmonium modes is of special interest and
  further provides complementary information on   physics beyond the SM.
In this paper, we first  calculated the $B_c\rightarrow \chi_{c0},\chi_{c1},\chi_{c2},h_c$ transition
form factors at the small momentum region within the improved PQCD framework.
By fitting an auxiliary three-parameter exponential function we obtained the momentum-squared-dependent 
form factors in the full kinematical region.
 We used them to estimate 
 the branching ratios of the considered semileptonic. The order of branching ratios shows
that these channels are accessible in the near future experiments.
 We also gave predictions on the ratio between the tau and light lepton branching ratio $\mathcal {R}(X)$,
  which are smaller than our previous calculation of $\mathcal {R}(J/\psi)$ due to the suppression from the phase space.
  A future improvable measurement might reveal whether a similar anomaly  exists in these ratios.
  Three  polarization contributions were also investigated in detail for the axial-vector and  tensor modes.
  The approximately equal polarization fractions between the tau and   light lepton with the same charmonium in the final states
  may indicate that the lepton flavor universality violation is negligible
  in  the relative polarization contributions. 
These results and findings will be further tested by the LHCb and Belle II experiments in the near future.
\begin{acknowledgments}
I would like to acknowledge Wei Wang, Hsiang-nan Li and Xin Liu for helpful discussions. This work is supported in
part by the National Natural Science Foundation of China
under Grants No. 11605060 and No. 11547020, in part
by the Program for the Top Young Innovative Talents of
Higher Learning Institutions of Hebei Educational
Committee under Grant No. BJ2016041, and in part by Training Foundation
of North China University of Science and Technology under Grants
No. GP201520 and No. JP201512.
\end{acknowledgments}

\end{document}